\documentclass[doublecol]{epl2}

\newcommand{\be}{\begin{equation}}
\newcommand{\ee}{\end{equation}}
\newcommand{\bea}{\begin{eqnarray}}
\newcommand{\eea}{\end{eqnarray}}

\title{Nonexistence of classical diamagnetism and nonequilibrium \\
fluctuation theorems for charged particles on a curved surface}
\shorttitle{Classical diamagnetism and the fluctuation theorems}

\author{Punyabrata Pradhan \and Udo Seifert}
\shortauthor{P. Pradhan and U. Seifert}

\institute{II. Institut f\"ur Theoretische Physik, Universit\"at Stuttgart,
Stuttgart 70550, Germany}


\abstract{We show that the classical Langevin dynamics for a
charged particle on a closed curved surface in a time-independent
magnetic field leads to the canonical distribution in the long
time limit. Thus the Bohr-van  Leeuwen theorem holds even for a
finite system {\it without any boundary} and the average magnetic
moment is zero. This is contrary to the recent claim by Kumar and
Kumar (EPL, {\bf 86} (2009) 17001), obtained from numerical
analysis of Langevin dynamics, that a classical charged particle
on the surface of a sphere in the presence of a magnetic field has
a nonzero average diamagnetic moment. We extend our analysis to a
many-particle system on a curved surface and show that the
nonequilibrium fluctuation theorems also hold in this geometry.}

\begin{document}

\maketitle

{\it Introduction} - The Bohr-van Leeuwen (BvL) theorem states
that the average magnetic moment of a classical system consisting
of charged particles in the presence of an external
time-independent magnetic field is zero in thermal equilibrium
\cite{vanVleck}. The proof is in principle simple: Since the free
energy, calculated from the canonical partition function, is
independent of the external magnetic field, the average magnetic
moment, which is the derivative of the free energy with respect to
the magnetic field, is therefore identically zero.

At first glance the statement of the BvL theorem may appear
disconcerting and counter-intuitive as, in the presence of an
external magnetic field, the charged particles undergo orbital
motion. Thus, one could expect the system to have an average
nonzero magnetic moment. But according to the BvL theorem, this is
not the case. To understand the null classical magnetic moment
physically, it is often pointed out that the boundary of a system
plays a subtle role \cite{vanVleck, Jayannavar_Kumar, Kumar_Kumar,
Dattagupta_Singh}. It is argued that the charged particles in the
bulk undergo orbital motion which gives rise to a nonzero
diamagnetic moment, but there is also a paramagnetic moment
arising due to incomplete orbits of particles which bounce off the
boundary in a cuspidal manner. This paramagnetic contribution
exactly cancels the diamagnetic one so that the net magnetic
moment is zero. Recently, based on this intuitive picture, Kumar
and Kumar have considered a finite system in the presence of a
magnetic field \cite{Kumar_Kumar} where there is no boundary, such
as a case of a particle moving on the surface of a sphere. These
authors have claimed, from a numerical analysis of Langevin
dynamics, that there exists a nonzero classical diamagnetic
moment. It has been argued that the nonzero magnetic moment arises
due to the avoided cancellation of the diamagnetic moment in the
bulk and the paramagnetic moment having no contribution due to the
absence of any boundary.

However, in this paper, we demonstrate that the above argument is
incorrect. To this end, we consider a particle moving on a closed
surface {\it without any boundary} in the presence of a constant
magnetic field. We show analytically that, in the long time limit,
the system which is governed by the classical Langevin dynamics is
indeed described by the equilibrium canonical distribution. The
case of a particle moving on a  sphere, considered by Kumar and
Kumar earlier \cite{Kumar_Kumar}, is one special case of a class
of systems we study here. Recently, Kaplan and Mahanti
\cite{Kaplan_Mahanti} have considered Langevin dynamics of a
particle constrained to move along a circle, but since in this
case the motion of the particle is unaffected by the magnetic
field, their analysis does not necessarily disprove the claim of
Kumar and Kumar in full generality. However, motion on a sphere
does depend on the magnetic field and therefore it is a first
nontrivial case which will be studied here. Our results show that
the previously claimed role of a boundary for having a zero
diamagnetic moment is a misleading one because the average
magnetic moment, as argued here, must vanish even for a finite
boundary-less system which is shown to have a canonical
distribution in thermal equilibrium. In principle, one can
understand the reason for vanishing of the magnetic moment as
following. In thermal equilibrium,  the probabilities for a
particle to have velocity $\vec{v}$ and $-\vec{v}$, at any given
position $\vec{r}$, are equal and
therefore the average magnetic moment $(q/2c) \langle \vec{r}
\times \vec{v}\rangle$ of a particle of charge $q$ is zero.

We also generalize our results to a Newtonian many-particle system
with charged  particles moving on a curved surface of an arbitrary
shape in the presence of a time-independent magnetic field. We
show that Liouville's theorem holds in the absence of any thermal
noise and damping. By adding those, the thermal distribution
becomes the usual canonical one. At the end, we demonstrate with
the Jarzynski equality \cite{JarzynskiPRL1997} and the Crooks
theorem \cite{CrooksPRE1999} that paradigmatic cases of the
nonequilibrium fluctuation theorems (reviewed in \cite{review1,
review2, Seifert}) are valid even for a system with particles on a
curved surface in the presence of an external time-dependent
magnetic field and other nonconservative forces. The fluctuation
theorems in the presence of a time-dependent magnetic field in
Euclidean space have been studied in \cite{Jayannavar, Pradhan}.

{\it Newtonian dynamics of a single particle} - We start in the
simplest setting where we first consider deterministic motion of a
single particle of charge $q$ and mass $m$ moving in the presence
of a uniform time-independent external magnetic field $\vec{B}$ applied
along the $z$-axis. Newton's equation of motion can be simply
written as, \be m \dot{\vec{v}} = \frac{q}{c} (\vec{v} \times \vec{B})
\label{Newton0} \ee where $\vec{v}$ the velocity of the particle and
$\vec{B}=B \hat{z}$, $\hat{z}$ being the unit vector along
the $z$-axis. In the presence of the constraint that the particle moves
on the surface of a sphere with radius $a$, it is convenient to
switch to spherical coordinate $\{ r, \theta, \phi \}$
where the respective unit vectors are denoted as $\hat{r}$,
$\hat{\theta}$, $\hat{\phi}$ and the constraint is expressed as
$r=constant=a$. By equating force and acceleration terms along
$\hat{\theta}$ and $\hat{\phi}$ directions, Eq. (\ref{Newton0})
can be rewritten as \bea m a (\ddot{\theta} - \dot{\phi}^2 \sin
\theta \cos \theta) \hat{\theta} = (\frac{qB}{c} a \dot{\phi} \sin
\theta \cos \theta) \hat{\theta} \mbox{~}, \label{Newton1} \\ m a
(\ddot{\phi} \sin \theta + 2 \dot{\theta} \dot{\phi} \cos \theta)
\hat{\phi} = - (\frac{qB}{c}a \dot{\theta} \cos \theta) \hat{\phi}
\label{Newton2}. \eea We convert these two second-order
differential equations to four first-order differential equations
in the $4$-dimensional phase space $\{ \theta, \phi, v_\theta,
v_\phi \}$, \bea \dot{\theta}=\frac{v_{\theta}}{a} \equiv
{\cal{V}}_{\theta} \label{Newton_phase_space1} \mbox{~},\\
\dot{\phi} = \frac{v_{\phi}}{a \sin \theta} \equiv
{\cal{V}}_{\phi} \mbox{~}, \label{Newton_phase_space2} \\
\dot{v}_{\theta} = a \left[ \frac{\cos \theta}{a^2 \sin \theta}
v_{\phi}^2 + \frac{\omega_c \cos \theta}{a} v_{\phi} \right]
\equiv {\cal{V}}_{v_{\theta}}  \mbox{~},
\label{Newton_phase_space3} \\ \dot{v}_{\phi} = - a \left[
\frac{\cos \theta}{a^2 \sin \theta} v_{\theta} v_{\phi} +
\frac{\omega_c \cos \theta}{a} v_{\theta} \right] \equiv
{\cal{V}}_{v_{\phi}} \mbox{~}, \label{Newton_phase_space4} \eea using the
relations $\ddot{\theta} = \dot{v}_{\theta}/a$ and $\ddot{\phi} =
(\dot{v}_{\phi} - a \dot{\theta} \dot{\phi} \cos \theta)/(a \sin
\theta)$ respectively in Eqs. (\ref{Newton1}) and (\ref{Newton2}),
and writing the cyclotron frequency $w_c = qB/mc$.

The time evolution of the phase space density $\rho$, at a phase
space point $\{ \theta, \phi, v_{\theta}, v_{\phi} \}$, is
governed by Liouville's equation which can be obtained from the
equation of continuity in the phase space, \be \frac{\partial
\rho}{\partial t} + \frac{\partial (\rho
{\cal{V}}_{\theta})}{\partial \theta} + \frac{\partial (\rho
{\cal{V}}_{\phi})}{\partial \phi} + \frac{\partial (\rho
{\cal{V}}_{v_{\theta}})}{\partial v_{\theta}} + \frac{\partial
(\rho {\cal{V}}_{v_{\phi}})}{\partial v_{\phi}} = 0 \label{L1}.
\ee The time-independent solution of Liouville's equation,
${\partial \rho_{\rm{st}}}/{\partial t}=0$, will be required later
when we consider stochastic Langevin dynamics and deterministic
many-particle Newtonian dynamics of a system in the presence of a
magnetic field. Since ${\partial {\cal V}_\theta}/{\partial
\theta} = {\partial {\cal V}_\phi}/{\partial \phi} = {\partial
{\cal V}_{v_\theta}}/{\partial v_{\theta}} = 0$ and ${\partial
{\cal V}_{v_{\phi}}}/{\partial v_{\phi}} = - ({\cos \theta}/{a
\sin \theta}) v_{\theta}$, the time-independent solution
$\rho_{\rm{st}}$ is simply given by \be \rho_{\rm{st}} =
\rho_{\rm{mic}} \sin \theta = \frac{\sqrt{m}}{8 \pi^2 a^2
\sqrt{2E}} \sin \theta, \label{Liouville_solution1} \ee where the
numerical value of the microcanonical probability distribution
$\rho_{\rm{mic}}$ follows from the normalization condition \bea
\int d \Lambda(\theta, \phi, v_{\theta}, v_{\phi}) \rho_{\rm{mic}}
\delta(m(v_{\theta}^2 + v_{\phi}^2)/2 - E) = 1 \label{vol_measure}
\eea with $\int d \Lambda = a^2 \int_{0}^{\pi} \sin \theta d\theta
\int_{0}^{2\pi} d\phi \int_{-\infty}^{\infty} dv_{\theta}
\int_{-\infty}^{\infty} dv_{\phi}$ denoting integration over
volume in the phase space $\{ \theta, \phi, v_{\theta}, v_{\phi}
\}$ and $E$ the total kinetic energy of the particle. Clearly the
factor of $\sin \theta$ in Eq. (\ref{Liouville_solution1}) comes
from the measure of an infinitesimal area on the surface of a
sphere in  spherical coordinate. The statement of Liouville's
theorem that the phase space is incompressible under Newtonian
time-evolution can now be expressed as $d \tilde{\rho}/dt = 0$
where $\tilde{\rho} = (\rho/\sin \theta)$. Note that the
time-independent magnetic field does not perform any work (nor do
the constraint-forces) on the particle, implying ${\rm d}{
E}/{\rm d}t=m(v_{\theta} \dot{v}_{\theta} + v_{\phi} \dot{v}_{\phi})=0$,
and using Eqs. (\ref{Newton_phase_space3}) and
(\ref{Newton_phase_space4}), we get \be  (v_{\theta} {\cal
V}_{v_{\theta}} + v_{\phi} {\cal V}_{v_{\phi}})=0.
\label{Constant_E}\ee

{\it Langevin dynamics} - Now we analyse the problem considered in
Ref. \cite{Kumar_Kumar}, i.e., the classical Langevin dynamics of
a charged particle constrained to move on the surface of a sphere
in the presence of an external uniform magnetic field. The
Langevin equations are obtained by adding a noise term and a
friction force to Newton's equations, Eqs.
(\ref{Newton_phase_space3}) and (\ref{Newton_phase_space4}), \bea
\dot{v}_{\theta} = {\cal V}_{v_{\theta}} - \gamma v_{\theta} +
\sqrt{{2 \gamma k_B T}/{m}} f_{\theta} \mbox{~}, \\ \dot{v}_{\phi}
= {\cal V}_{v_{\phi}} - \gamma v_{\phi} + \sqrt{{2 \gamma k_B
T}/{m}} f_{\phi} \mbox{~}, \eea in addition to Eqs.
(\ref{Newton_phase_space1}) and (\ref{Newton_phase_space2}), where
$\gamma$ is the friction coefficient, $k_B$ is the Boltzmann
constant, $T$ is the temperature, $\vec{f}$ is the Gaussian
white-noise satisfying the fluctuation-dissipation theorem,
$\langle f_{\alpha}(t) f_{\alpha'}(t') \rangle = \delta_{\alpha
\alpha'} \delta(t-t')$ and the subscript corresponds to the
component along unit vector $\hat{\alpha}$ or $\hat{\alpha}'$ with
$\alpha, \alpha' \equiv \theta, \phi$. The probability density
$P(\theta, \phi, v_{\theta}, v_{\phi})$ is governed by the
Fokker-Planck equation \cite{Risken}, \bea \frac{\partial
P}{\partial t} = -\frac{\partial ({{\cal V}}_{\theta} P)}{\partial
\theta} -\frac{\partial ({{\cal V}}_{\phi} P)}{\partial \phi}
-\frac{\partial ({{\cal V}}_{v_{\theta}} P)}{\partial v_{\theta}}
- \frac{\partial ({{\cal V}}_{v_{\phi}} P)}{\partial v_{\phi}}
\nonumber \\ + \frac{\partial (\gamma v_{\theta} P)}{\partial
v_{\theta}} + \frac{\partial (\gamma v_{\phi} P)}{\partial
v_{\phi}} + \frac{\gamma k_B T}{m} \left[ \frac{\partial^2
P}{\partial v_{\theta}^2} + \frac{\partial^2 P}{\partial
v_{\phi}^2} \right]. \mbox{~} \label{FP} \eea To solve the steady
state of the above equation, i.e. the solution for which
${\partial P_{\rm{st}}}/{\partial t}=0$, we try the equilibrium
ansatz $P_{\rm{st}}(\theta, \phi, v_{\theta}, v_{\phi}) = const
\times \rho_{\rm{st}} \times P_1(v_{\theta}, v_{\phi})$ where
$\rho_{\rm{st}}$ is the time-independent solution of Eq.
(\ref{L1}) and $P_1(v_{\theta}, v_{\phi}) =
\exp[-m(v_{\theta}^2+v_{\phi}^2)/2k_B T]$. Using ${\partial
P_1}/{\partial v_{\theta}} = - ({m}/{k_B T}) v_{\theta} P_1$,
${\partial P_1}/{\partial v_{\phi}}=- (m/k_B T) v_{\phi} P_1$ and
Eq. (\ref{Constant_E}), the steady-state probability distribution
for a particle moving on the surface of a sphere turns out to be
the equilibrium canonical probability distribution function, \bea
\rho_{\rm{can}} (\theta, \phi, v_{\theta}, v_{\phi}) =
\frac{P_{\rm{st}}}{\sin \theta} = \frac{m}{8 \pi^2 a^2 k_B T}
e^{-{\cal E}/k_B T}, \label{can_distribution} \eea where ${\cal E}
= m (v_{\theta}^2+v_{\phi}^2)/2$ and the normalization condition
being $\int d\Lambda \rho_{\rm{can}} = 1$. The probability of
finding the particle at any point of the sphere is uniform.
According to the general theory of Fokker-Planck equations
\cite{Risken}, any initial distribution will relax to a unique
stationary distribution, i.e., in this case the canonical one, in
the long time limit. Thus one can see that the average magnetic
moment is zero since $\langle v_{\theta} \rangle = \langle
v_{\phi} \rangle = 0$ with this distribution, and hence there is
no classical diamagnetism.

{\it Langevin dynamics with an external potential} - The above
analysis can be easily extended to the case when there is an
external potential $U(\vec{r})$ at a position $\vec{r}$ due to
conservative forces in addition to an external magnetic field. The
Langevin equation is then written as \be m \dot{\vec{v}} = q
(\vec{v} \times \vec{B})/c - \vec{\nabla} U - \gamma m \vec{v} +
\sqrt{2 \gamma m k_B T} \vec{f}, \label{Langevin_potential} \ee
where $\vec{f}$ is the Gaussian white-noise satisfying the
fluctuation-dissipation theorem. When the particle is constrained
to move on the surface of a sphere, Eq. (\ref{Langevin_potential})
can be rewritten by equating respective components of forces and
accelerations along $\hat{\theta}$ and $\hat{\phi}$ directions,
\bea \dot{v}_{\theta} = {\cal V}_{v_{\theta}} - (\vec{\nabla}
U)_\theta/m - \gamma v_\theta + \sqrt{{2 \gamma k_B T}/{m}}
f_{\theta}\mbox{~}, \label{L_sphere_pot1} \\
\dot{v}_{\phi} = {\cal V}_{v_{\phi}} -  (\vec{\nabla} U)_\phi/m -
\gamma v_\phi + \sqrt{{2 \gamma k_B T}/{m}} f_{\phi}\mbox{~},
\label{L_sphere_pot2} \eea where the gradient $\vec{\nabla}
U(\vec{r})$ is expressed in spherical coordinate with
$(\vec{\nabla} U)_\theta = (1/a) (\partial U/\partial \theta)$ and
$(\vec{\nabla} U)_\phi = (1/a \sin \theta) (\partial U/\partial
\phi)$ respectively being the $\hat{\theta}$ and $\hat{\phi}$
components of the external conservative-forces. The full set of
Langevin equations now consists of Eqs.
(\ref{Newton_phase_space1}), (\ref{Newton_phase_space2}),
(\ref{L_sphere_pot1}) and (\ref{L_sphere_pot2}). The corresponding
Fokker-Planck equation can be written as \bea \frac{\partial
P}{\partial t} = - \frac{\partial ({\cal V}_\theta P)} {\partial
\theta} - \frac{\partial ({\cal V}_\phi P)} {\partial \phi} -
\frac{\partial ({\cal V}_{v_\theta} P)} {\partial {v_\theta}} -
\frac{\partial ({\cal V}_{v_\phi} P)} {\partial {v_\phi}} \nonumber \\
+ \frac{\partial \left(\frac{1}{m} (\vec{\nabla} U)_\theta P
\right)}{\partial v_\theta} + \frac{\partial \left(\frac{1}{m}
(\vec{\nabla} U)_\phi P \right)} {\partial v_\phi} \nonumber \\ +
\frac{\partial (\gamma v_{\theta} P)}{\partial v_{\theta}} +
\frac{\partial (\gamma v_{\phi} P)}{\partial v_{\phi}} +
\frac{\gamma k_B T}{m} \left[ \frac{\partial^2 P}{\partial
v_{\theta}^2} + \frac{\partial^2 P}{\partial v_{\phi}^2} \right].
\mbox{~} \label{FP_potential} \eea It is straightforward to show
that the time-independent solution of the Fokker-Planck equation,
$\partial P_{\rm{st}}/\partial t =0$, can be written as $P_{\rm
st} = (\sin \theta) \rho_{\rm can}$ where the canonical
probability distribution $\rho_{\rm can}$ is given by \bea
\rho_{\rm{can}}(\theta, \phi, v_\theta, v_\phi) = ( {e^{-{\cal E}/
k_B T}} )/{Z} \eea with the total energy ${\cal E} = [ m
(v^2_\theta+v^2_\phi)/2 + U(\theta, \phi) ]$ and the partition
function $Z = \int d\Lambda \exp(-{\cal E}/k_B T)$. Clearly, the
probability distribution, and therefore the partition function,
are both independent of the magnetic field.

{\it Newtonian dynamics of a many-particle system} - Now we
consider a classical Newtonian system consisting of $N$
interacting charged particles, with mass $m$ and charge $q$,
moving on the surface of a sphere in the presence of a constant
external magnetic field $\vec{B}(\vec{r})$ which may be
space-dependent. Newton's equation of motion of the $i$-th
particle can be written as \be \left( m \dot{\vec{v}}_i
\right)_{\hat{\alpha}} = \left( \frac{q}{c} (\vec{v}_i \times
\vec{B}(\vec{r}_i)) \right)_{\hat{\alpha}} - \left(
\vec{\nabla}_{\vec{r}_i} U(\{\vec{r}_i\}) \right)_{\hat{\alpha}}
\ee where $\vec{r}_i$, $\vec{v}_i$ position and velocity of the
$i$-the particle, $(\dots)_{\hat{\alpha}}$ denotes a component
along the unit vector $\hat{\alpha}$ with $\alpha \equiv \theta,
\phi$, and $U(\{\vec{r}_i\})$ is the scalar potential containing
both internal pair-potentials as well as an external potential.
Similarly as in the previous single-particle case of Eqs.
(\ref{Newton_phase_space1})-(\ref{Newton_phase_space4}), Newton's
equations of motion can be written in spherical coordinate in
$4N$-dimensional phase space $\{\theta_i, \phi_i, v_{\theta_i},
v_{\phi_i} \}$ as \bea \dot{\theta}_i=\frac{v_{\theta_i}}{a}
\equiv {\cal{V}}^{(i)}_{\theta},
\label{Newton_many_partile_phase_space1}\\
\dot{\phi}_i = \frac{v_{\phi_i}}{a \sin \theta_i} \equiv
{\cal{V}}_{\phi}^{(i)}, \label{Newton_many_partile_phase_space2} \\
\dot{v}_{\theta_i} =  {\cal{V}}_{v_{\theta}}^{(i)} - \frac{1}{m}
\left( \vec{\nabla}_{\vec{r}_i} U(\{\vec{r}_i\})
\right)_{\hat{\theta}_i},
\label{Newton_many_partile_phase_space3} \\
\dot{v}_{{\phi}_i} = {\cal{V}}_{v_{\phi}}^{(i)} - \frac{1}{m}
\left( \vec{\nabla}_{\vec{r}_i} U(\{\vec{r}_i\})
\right)_{\hat{\phi}_i}, \label{Newton_many_partile_phase_space4}
\eea where ${{\cal V}}_{v_{\theta}}^{(i)} = a [ ({\cos
\theta_i}/{a^2 \sin \theta_i} ) v_{\phi_i}^2 + {(\omega_c \cos
\theta_i}/{a}) v_{\phi_i} ]$ and $\left( \vec{\nabla}_{\vec{r}_i}
U(\{\vec{r}_i\}) \right)_{\hat{\theta}_i} = (1/a) (\partial
U(\{\theta_i, \phi_i\})/\partial \theta_i)$,
${\cal{V}}_{v_{\phi}}^{(i)} = - a \left[ ({\cos \theta_i}/{a^2
\sin \theta_i}) v_{\theta_i} v_{\phi_i} + ({\omega_c \cos
\theta_i}/{a}) v_{\theta_i} \right]$ and $\left(
\vec{\nabla}_{\vec{r}_i} U(\{\vec{r}_i\}) \right)_{\hat{\phi}_i}
= (1/a \sin \theta_i ) (\partial U(\{\theta_i, \phi_i\})/\partial
\phi_i)$. The time evolution of the phase space density
$\rho(\{\theta_i, \phi_i, v_{\theta_i}, v_{\phi_i}\})$ in
$4N$-dimensional phase space is given by the continuity equation,
\bea \frac{\partial \rho}{\partial t} + \sum_{i=1}^{N} \left[
\frac{\partial (\rho {\cal V}^{(i)}_\theta)}{\partial \theta_i} +
\frac{\partial (\rho {\cal V}^{(i)}_\phi)}{\partial \phi_i} +
\frac{\partial (\rho {\cal V}^{(i)}_{v_\theta})}{\partial
v_{\theta_i}} + \frac{\partial (\rho {\cal
V}^{(i)}_{v_\phi})}{\partial v_{\phi_i}} \right] \nonumber \\ = 0.
\label {L_many_particle} \eea Using ${\partial {\cal
V}^{(i)}_\theta}/{\partial \theta_i} = {\partial {\cal
V}^{(i)}_\phi}/{\partial \phi_i} = {\partial {\cal
V}^{(i)}_{v_\theta}}/{\partial v_{\theta_i}} = 0$ and ${\partial
{\cal V}^{(i)}_{v_\phi}}/{\partial v_{\phi_i}} = - ({\cos
\theta_i}/{a \sin \theta_i}) v_{\theta_i}$, it is straightforward
to see that the time-independent solution of Eq.
(\ref{L_many_particle}), $\partial \rho_{\rm{st}}/\partial t = 0$,
is given by \be \rho_{\rm{st}}(\{\theta_i, \phi_i, v_{\theta_i},
v_{\phi_i}\}) = \rho_{\rm{mic}} \left( \prod_{i=1}^{N} \sin
\theta_i \right). \label{microcan_sphere} \ee The microcanonical
probability measure is $ \rho_{\rm{mic}} = {1}/{\Omega(E)} =
const$ and $\Omega(E)$ is the total phase volume of a constant
energy surface in $4N$-dimensional phase space i.e., $\Omega(E) =
\prod_i \left( \int d\Lambda(\theta_i, \phi_i, v_{\theta_i},
v_{\phi_i}) \right) \delta({\cal E}(\{\theta_i, \phi_i,
v_{\theta_i}, v_{\phi_i}\}) - E)$ where $d\Lambda$ is the
infinitesimal volume-measure as defined below Eq.
(\ref{vol_measure}) and the total energy of the system is ${\cal
E} = [(m/2) \sum_i (v_{\theta_i}^2 + v_{\phi_i}^2) + U(\{\theta_i,
\phi_i\}) ]$. Also note that $d \tilde{\rho}/dt=0$ with
$\tilde{\rho} = \rho/(\prod_i \sin \theta_i)$, which is the
statement of Liouville's theorem. Therefore, assuming ergodicity,
an isolated system in equilibrium can be described by a
microcanonical distribution where any time-averages can be in
principle calculated from the uniform probability distribution on a
constant energy surface. Since, in this case, two configurations
with the same position coordinates and opposite velocities, i.e,
$\{\theta_i, \phi_i, v_{\theta_i}, v_{\phi_i} \}$ and $\{\theta_i,
\phi_i, -v_{\theta_i}, -v_{\phi_i} \}$, are equally probable,  the
ensemble (microcanonical) average of the total magnetic moment $
\sum_i (q/2c) \langle (\vec{r}_i \times \vec{v}_i) \rangle$ is
identically zero.

{\it Motion on an arbitrary curved surface} - We now consider
dynamics of a particle of mass $m$ and charge $q$ which is
constrained to move on a closed surface of an arbitrary shape in the
presence of an externally applied constant uniform magnetic field
$\vec{B}=B \hat{z}$. The position of the particle in a
three-dimensional space is specified by a set of general
orthogonal curvilinear coordinates $\{q^{\mu}\}$ where $\mu=1,2,3$
and the respective unit vector is denoted as $\hat{q}_{\mu}$. The
constraint that the particle moves on a surface can be imposed,
without loss of any generality, by taking $q^1 = constant$ or
equivalently $\dot{q}^1=0$. Then a spatial point on the surface
$q^1=constant$ is specified by the remaining two generalized
coordinates, i.e., $\{q^2, q^3\}$. An infinitesimal line element
$dl$ in the curvilinear coordinate can be written as \be (dl)^2=
\sum_{\mu, \nu=1}^3 g_{\mu \nu} dq^{\mu} dq^{\nu}, \ee where
$g_{\mu \nu}$ is the metric tensor of the curved surface,
represented as a $3 \times 3$ matrix. Since the coordinate system
is orthogonal, the metric tensor is diagonal, i.e., $g_{\mu \nu} =
(h_{\mu})^2 \delta_{\mu \nu}$ where $h_{\mu}$ is called the scale
factor associated with the curvilinear coordinate $q^{\mu}$ and
$\delta_{\mu \nu}$ is the Kronecker-delta function. For example,
in spherical coordinate considered previously, $q^1=r$,
$q^2=\theta$ and $q^3=\phi$ and the respective scale factors are
$h_{r}=1$, $h_{\theta}=r$ and $h_{\phi}=r \sin \theta$. We also
define a matrix element $g^{\mu \nu}$ which is inverse of the $3
\times 3$ matrix representing the metric tensor $g_{\mu \nu}$,
i.e., $\sum_{\delta=1}^3 g^{\mu \delta} g_{\delta \nu}=\delta_{\mu
\nu}$. Since the metric tensor is diagonal, the inverse is also
diagonal, i.e., $g^{\mu \nu} = (h_{\mu})^{-2} \delta_{\mu \nu}$.
An infinitesimal volume element $dV$ and surface element $dA$ of
the surface $q^1=constant$ can be written as $dV=h_1 h_2 h_3 dq^1
dq^2 dq^3$ and $dA = h_2 h_3 dq^2 dq^3$.

First we consider the deterministic motion of a single particle.
Newton's equation of motion can be written in the {\it covariant}
form \cite{Weinberg}, \be \ddot{q}^{\lambda} + \sum_{\mu, \nu=1}^3
\Gamma_{\mu \nu}^{\lambda} \dot{q}^{\mu} \dot{q}^{\nu} =
\frac{F^{\lambda}}{h_{\lambda}}, \label{Newton_curved1} \ee where
$F^{\lambda}$ is the component of an external force along unit
vector $\hat{q}^{\lambda}$ and the {\it affine connection}
$\Gamma_{\mu \nu}^{\lambda}$ in an orthogonal curvilinear
coordinate with $g_{\mu \nu} = \delta_{\mu \nu} g_{\mu \mu}$ can
be written as \be \Gamma_{\mu \nu}^{\lambda} = \frac{1}{2}
g^{\lambda \lambda} \left[ \delta_{\nu \lambda} \frac{\partial
g_{\lambda \lambda}}{\partial q^{\mu}} + \delta_{\mu \lambda}
\frac{\partial g_{\lambda \lambda}}{\partial q^{\nu}} -
\delta_{\nu \mu} \frac{\partial g_{\mu \mu}}{\partial q^{\lambda}}
\right]. \label{affine_connection2} \ee Note that we do not follow
the Einstein summation convention here. Now defining the velocity
$v^{\lambda}$ along an unit vector $\hat{q}^{\lambda}$ as
$v^{\lambda} = h_{\lambda} \dot{q}^{\lambda}$ and using the
relations $\ddot{q}^{\lambda} = (\dot{v}^{\lambda}-
\dot{h}_{\lambda} \dot{q}^{\lambda})/h_{\lambda}$,
$\dot{h}_{\lambda} = \sum_{\mu} (\partial h_{\lambda}/\partial
q^{\mu}) \dot{q}^{\mu}$ in Eq. (\ref{Newton_curved1}), we get a
set of two first-order differential equations, \bea
\dot{q}^{\lambda} = \frac{v^{\lambda}}{h_{\lambda}} \equiv {\cal
V}_{q^{\lambda}}\mbox{~}, \label{q_dot} \\ \dot{v}^{\lambda} =
\sum_{\mu=2}^{3} \frac{\partial h_{\lambda}}{\partial q^{\mu}}
\frac{v^{\mu}}{h_{\mu}} \frac{v^{\lambda}}{h_{\lambda}} -
h_{\lambda} \sum_{\mu, \nu = 2}^3 \Gamma_{\mu \nu}^{\lambda}
\frac{v^{\mu}}{h_{\mu}} \frac{v^{\nu}}{h_{\nu}} \nonumber \\ +
\frac{1}{m} F_{\rm mag}^{\lambda} \equiv {\cal
V}_{v^{\lambda}}\mbox{~}, \label{v_dot} \eea where
$F^{\lambda}_{\rm{mag}}$ is the component of the Lorentz force
along the unit vector $\hat{q}^{\lambda}$ acting on the particle
in the presence of a magnetic field. Note that the indices
$\lambda$, $\mu$ and $\nu$ are specified to take values $2$ and
$3$ only as the motion along unit vector $\hat{q}^1$ is not
possible, i.e., $v^1=0$. The Lorentz force is given by
$\vec{F}_{\rm{mag}} = (q/c) (\vec{v} \times \vec{B})$ where
$\vec{v}\times \vec{B} = (v^2 \hat{q}^2 + v^3 \hat{q}^3) \times B
\hat{z}$. Now taking the components of the Lorentz force tangential
to the surface, we get $F_{\rm{mag}}^2 = (qB/c) v^3 \cos \Theta$
and $F_{\rm{mag}}^3 = -(qB/c) v^2 \cos \Theta$ where $\cos \Theta=
(\hat{q}^1 \stackrel{.}{} \hat{z})$, $\Theta$ being the angle
between unit vectors $\hat{q}^1$ and $\hat{z}$. The angle $\Theta$
is a function of generalized coordinates $\{q^{\mu}\}$, i.e.,
$\Theta = \Theta(\{q^{\mu}\})$. For example in spherical
coordinate, using explicit expressions of the scale factor
$h_{\lambda}$ and the affine connection $\Gamma_{\mu
\nu}^{\lambda}$, one could check that Newton's equations of motion
in Eq. (\ref{q_dot}) and (\ref{v_dot}) reproduce Eqs.
(\ref{Newton_phase_space1})-(\ref{Newton_phase_space4}) where in
this case $\Theta = \theta$. Since the external magnetic force and
the constraint-forces do not perform work on the particle, using
Eq. (\ref{v_dot}), one can check that  \be \sum_{\lambda=2}^3
\dot{v}^{\lambda} v^{\lambda}= 0 = \sum_{\lambda=2}^3 {\cal
V}_{v^{\lambda}} v^{\lambda}\mbox{~},
\label{energy_conservation_curved} \ee implying the total kinetic
energy energy ${\cal E} = (m/2) \sum_{\lambda=2}^3 (v^{\lambda})^2
= const$ in this case.

Alternatively, to derive Newton's equations of motion of a
particle constrained to move on a curved surface, one can as well
start with the Lagrangian $L = K(\{q^{\lambda}, \dot{q}^{\lambda}\})
- U(\{q^{\lambda}\})$ where $K =(m/2) \sum_{\lambda} (v^{\lambda})^2$
is the total kinetic energy and $U$ is the total potential energy
where the equation of motion is given by  $d ( {\partial
L}/{\partial \dot{q}^{\lambda}})/dt - {\partial L}/{\partial
q^{\lambda}}  = 0$  \cite{Casey}. Using the affine
connection in Eq. (\ref{affine_connection2}), one can check
that Newton's equation in Eq. (\ref{Newton_curved1}) is indeed
same as the equation derived from the Lagrangian.

The time evolution of the phase space density $\rho(\{q^{\lambda},
v^{\lambda}\})$ in 4-dimensional phase space $\{q^2, q^3, v^2,
v^3\}$ is governed by Liouville's equation, \be \frac{\partial
\rho}{\partial t} + \sum_{\lambda=2}^3 \left[ \frac{\partial (\rho
{\cal V}_{q^{\lambda}})}{\partial q^{\lambda}} + \frac{\partial
(\rho {\cal V}_{v^{\lambda}})}{\partial v^{\lambda}} \right] =0.
\label{LiouvilleCurvedTimeDependent1} \ee A straightforward
calculation shows that the time-independent solution of Eq.
(\ref{LiouvilleCurvedTimeDependent1})  is given by \be
\rho_{\rm{st}} (q^2, q^3, v^2, v^3) = \rho_{\rm{mic}} h_2 h_3
\mbox{~}, \ee where the microcanonical probability distribution
$\rho_{\rm{mic}} = {1}/{\Omega(E)} = const$ and the phase space
volume $\Omega(E)$ of a constant energy surface is given by
$\Omega(E) = (\int d\Lambda) \delta({\cal E}- E)$ with $d\Lambda =
\prod_{\lambda=2}^3 h_{\lambda} dq^{\lambda} dv^{\lambda}$ the
infinitesimal phase space volume-measure and ${\cal E} = (m/2)
\left[ (v^2)^2+(v^3)^2 \right]$ the total kinetic energy.
Liouville's equation can be cast in a manifestly covariant form in
terms of a scalar density variable $\tilde{\rho}=\rho/\sqrt{g}$
where $g$ is determinant of the metric tensor \cite{Tuckerman}.
Then Liouville's theorem of incompressibility of the phase space
can be shown to have a form $d \tilde{\rho}/dt=0$. Note that
Liouville's theorem holds even if one adds a position-dependent
{\it nonconservative force} on the right hand side of Eq.
(\ref{v_dot}), since $\partial \dot{v}^{\lambda}/\partial
v^{\lambda} = 0$ in this case. This will be required later to
prove the nonequilibrium fluctuation theorems.

The Langevin equation in the presence of additional external
conservative forces, derivable from a scalar potential
$U(\{q^{\lambda}\})$, can be written as \bea \dot{q}^{\lambda} = {\cal
V} _{q^{\lambda}}\mbox{~}, \label{Langevin_potential_curved1} \\
\dot{v}^{\lambda} = {\cal V}_{v^{\lambda}} -
(\vec{\nabla} U)_{q^{\lambda}}/m - \gamma v^{\lambda} +
\sqrt{{2 \gamma k_B T}/{m}} f_{q^{\lambda}}\mbox{~},
\label{Langevin_potential_curved2} \eea where $(\vec{\nabla}
U)_{q^{\lambda}} = (1/h_{\lambda}) (\partial U/\partial
q^{\lambda})$ is the component of the conservative forces along
the unit vector $\hat{q}^{\lambda}$ and $f_{q^{\lambda}}$ is a
white-noise satisfying $\langle f_{q^{\lambda}}(t)
f_{q^{\lambda'}}(t') \rangle = \delta_{\lambda \lambda'}
\delta(t-t')$. The Fokker-Planck equation, as in the case of a
single particle on a sphere in Eq. (\ref{FP_potential}), can be
written as \bea \frac{\partial P}{\partial t} = -
\sum_{\lambda=2}^3 \left[ \frac{\partial ({\cal V}_{q^{\lambda}}
P)} {\partial q^{\lambda}} + \frac{\partial ({\cal
V}_{v^{\lambda}} P)} {\partial v^{\lambda}} \right] \nonumber \\
+ \sum_{\lambda=2}^3 \frac{\partial (\frac{1}{m} (\vec{\nabla}
U)_{q^{\lambda}} P)}{\partial v^{\lambda}} + \sum_{\lambda=2}^3
\frac{\partial}{\partial v^{\lambda}} \left[ \gamma v^{\lambda} P
+ \frac{\gamma k_B T}{m} \frac{\partial P}{\partial v^{\lambda}}
\right]. \label{FP_curved_pot} \eea Proceeding as before, the
time-independent solution of Eq. (\ref{FP_curved_pot}) is given by
the equilibrium ansatz $P_{\rm{st}} = \rho_{\rm st} \rho_{\rm
can}$ with $\rho_{\rm st} = (h_2 h_3)$ the time-independent
solution of Eq. (\ref{LiouvilleCurvedTimeDependent1}) and the
canonical probability distribution
$\rho_{\rm{can}}(\{q^{\lambda},v^{\lambda}\})$, \be
\rho_{\rm{can}}(\{q^{\lambda},v^{\lambda}\}) = (e^{-{\cal E}/k_B
T})/{Z} \label{can_curved_pot1} \ee where ${\cal E} = [ (m/2)
\sum_{\lambda} (v^{\lambda})^2 + U(\{q^{\lambda}\}) ]$ and the
partition function $Z = \prod_{\lambda} (\int h_{\lambda}
dq^{\lambda} dv^{\lambda} ) \exp(-{\cal E}/k_B T)$. The average
magnetic moment can be shown to be zero since $\langle v^{\lambda}
\rangle=0$ with this canonical distribution. Note that here we
have considered under-damped Langevin dynamics on a curved
surface. For over-damped Langevin dynamics on a curved surface and
the corresponding covariant formulation of the Fokker-Planck
equation, see \cite{Graham, Christensen}.

In the many-particle case in a microcanonical set up, one can
generalize Eq. (\ref{microcan_sphere}) to a system consisting of
particles moving on a surface of an arbitrary shape and obtain the
microcanonical distribution, \be \rho_{\rm{mic}}
(\{q_i^{\lambda}, v_i^{\lambda}\}) = {1}/{\Omega(E)} = constant.
\label{microcan_curved_many_particle} \ee  The total phase space
volume $\Omega(E)$ of the constant energy surface in
$4N$-dimensional space is given by $\Omega(E) = \prod_{i, \lambda}
[ \int (h_{\lambda} {\rm d} q_i^{\lambda}) {\rm d} v_i^{\lambda}] 
\delta({\cal
E} - E)$, where $i$ is the particle index, ${\cal E}= [ (m/2)
\sum_{i, \lambda} (v_i^{\lambda})^2 + U(\{q_i^{\lambda},
v_i^{\lambda}\}) ]$ is the total energy of the many-particle
system. Since the total energy ${\cal E}$ is an even function of
velocity $v^{\lambda}$, two configurations, one with
$\{q_i^{\lambda}, v_i^{\lambda}\}$ and other with
$\{q_i^{\lambda}, - v_i^{\lambda}\}$, are equally probable, and
this implies that the average total magnetic moment is zero for
the microcanonical distribution. Adding dissipation and a noise
term satisfying the fluctuation-dissipation theorem to the
deterministic Newton's equation of motion of the $i$-th particle,
one obtains the Langevin equation which again can be shown leading
to the canonical distribution in the long time limit.

{\it The Jarzynski equality and the Crooks theorem} - We finally
discuss two remarkable relations called the Jarzynski equality
(JE) \cite{JarzynskiPRL1997} and the Crooks theorem (CT)
\cite{CrooksPRE1999}, which involve fluctuations of work done on a
system driven arbitrarily far away from equilibrium.

First we consider the Langevin dynamics of a particle, in the
presence of a time-dependent external magnetic field and
nonconservative forces $\vec{F}_{nc}$ which include a
nonconservative electric field $-\partial \vec{A}/\partial t$
induced by the time-varying magnetic field $\vec{B} = (\nabla
\times \vec{A})$, $\vec{A}$ being the vector potential. The
Langevin equation (\ref{Langevin_potential_curved1}) still holds
and, now adding the component of $\vec{F}_{nc}$ along unit vector
$\hat{q}^{\lambda}$ to Eq. (\ref{Langevin_potential_curved2}), we get
\bea \dot{v}^{\lambda} = {\cal V}_{v^{\lambda}} - \frac{1}{m}
(\vec{\nabla} U)_{q^{\lambda}}  + \frac{1}{m} F^{\lambda}_{nc}  -
\gamma v^{\lambda} + \sqrt{\frac{2 \gamma k_B T}{m}}
f_{q^{\lambda}}\mbox{~}. \label{Langevin_nc} \eea Let us consider
a process in a symmetric time interval $-{\cal T} < t < {\cal T}$
where $\vec{F}_{nc}$ is nonzero only in a subinterval $0 \le t \le
\tau$ with $\tau \ll {\cal T}$ so that, at $t = \pm {\cal T}$, the
system is described by the canonical distribution as in Eq.
(\ref{can_curved_pot1}). The work $W$ done by the external forces
and the heat $Q$ transferred from the heat bath to the system can
be written as \bea W = \int_{-\cal T}^{\cal T} \left( \sum_{\lambda}
F^{\lambda}_{nc} v^{\lambda}  +  \frac{\partial U}{\partial
\alpha} \dot{\alpha} \right) dt , \label{Work} \\ Q = \int_{-\cal T}^{\cal T}
m \left( -\gamma v^{\lambda} + \sqrt{2\gamma k_B T/m}
f_{q^{\lambda}} \right) v^{\lambda} dt, \label{Heat} \eea
where $\alpha(t)$ is an external time-dependent parameter in the potential
energy $U$ which varies only in the interval $0 \le t \le \tau$.
Under simultaneous reversal of time and the direction of the magnetic field,
the total work done by the external forces is odd, i.e.,
$W \rightarrow -W$ as $t \rightarrow -t$, $v^{\lambda} \rightarrow
- v^{\lambda}$ and $\vec{B} \rightarrow -\vec{B}$ (equivalently
$\vec{A} \rightarrow -\vec{A}$) since $F^{\lambda}_{nc}$, which includes
$-\partial \vec{A}/\partial t$, does not change sign. The
ratio of the probability ${\cal P}_F$ of a forward
path $\{q^{\lambda}(t), v^{\lambda}(t); \alpha(t), F^{\lambda}_{nc}(t)\}$ to the
probability ${\cal P} _R$ of a reverse path $\{ q^{\lambda}(-t),
-v^{\lambda}(-t); \alpha(-t), F^{\lambda}_{nc}(-t)\}$ is given by \be
\frac{{\cal P}_F}{{\cal P}_R} =  \exp \left[ -\beta \int_{-\cal T}^{\cal T}
\sum_{\lambda} {\cal F}^{\lambda} v^{\lambda} dt \right] \nonumber
= \exp(-\beta Q) \label{MR_canonical} \ee where ${\cal
F}^{\lambda} = [ m \dot{v}^{\lambda} + (\vec{\nabla}
U)_{q^{\lambda}} - F^{\lambda}_{nc} ]$, $\beta = 1/k_B T$ and we
have used Eqs. (\ref{energy_conservation_curved}),
(\ref{Langevin_nc}) and (\ref{Heat}). With the total energy of the
system ${\cal E} = [(m/2) \sum_{\lambda} (v^{\lambda})^2 + U ]$, and by
using Eqs. (\ref{energy_conservation_curved}) and
(\ref{Langevin_nc})-(\ref{Heat}), we get $\Delta {\cal E} = \int
(d{\cal E}/dt) dt = W + Q$ where $\Delta {\cal E}$ the change in the
total energy of the system. The probability distribution $P_F(W)$ of work
$W_F$ done for the forward protocol $\{\alpha(t), \vec{F}_{nc}(t) \}$ 
can be related to
the probability distribution $P_R(W)$ of work $W_R$ done for the
reverse protocol $\{\alpha(-t), \vec{F}_{nc}(-t)\}$ with the 
direction of $\vec{B}$
also reversed, \bea P_F(W) = \sum_{\mbox{forward paths}}
\rho_i  {\cal P}_F \delta(W_F - W) \nonumber
\\ = \sum_{\mbox{reverse paths}} \rho_f
{\cal P}_R  \delta(W_R + W) \times \left( e^{-\beta Q}
\frac{\rho_i}{\rho_f} \right) \nonumber \\ = e^{\beta(W-\Delta F)}
P_R(W), \label{CT} \eea where the subscripts $i$, $f$ denote the
corresponding initial and final quantity respectively. The
canonical probability distribution of the system $\rho =
\exp[-\beta({\cal E} - F)]$ is given in Eq.
(\ref{can_curved_pot1}) with $F=-(1/\beta) \ln Z$ being the free
energy, $\Delta F = F_f-F_i$, and we have also used Eq.
(\ref{MR_canonical}), ${\cal E}_f - {\cal E}_i = \Delta {\cal E} =
W_F + Q$ and $W_F=-W_R$. The relation in Eq. (\ref{CT}) is the
statement of the CT which has been shown here to be valid even for
a particle on a curved surface. The JE, $\langle \exp(-\beta W)
\rangle = \exp(-\beta \Delta F)$, can be obtained by integrating
the CT \cite{CrooksPRE1999}. It is straightforward to extend this
analysis to a many-particle system governed by Langevin dynamics
on a curved surface.

In a microcanonical set up, the JE and the CT have been recently
proved for a system consisting of particles moving in Euclidean
space in the presence of a time-dependent magnetic field and other
nonconservative forces. For particles moving on a curved surface,
the proof essentially follows from the fact that Liouville's
theorem still hold for the system and the heat bath combined in
the presence of a magnetic field and other nonconservative forces
\cite{Pradhan}.

{\it Summary} - We have shown that the classical Langevin dynamics
for charged particles on a closed curved surface in a time-independent
magnetic field leads to the canonical distribution in the long
time limit. Thus the Bohr-van  Leeuwen theorem holds even for a
finite system without any boundary and the average magnetic moment
for such classical systems is zero. Our analytical results
disproves the recent claim by Kumar and Kumar \cite{Kumar_Kumar}
that a classical charged particle on the surface of a sphere
governed by Langevin dynamics in the presence of a magnetic field
has a nonzero average diamagnetic moment. We also show that
nonequilibrium fluctuation theorems hold for a system consisting
of particles on a curved surface in the presence of a
time-dependent magnetic field and other nonconservative forces.

\acknowledgments

The authors thank K. Vijay Kumar for discussions and correspondence.
Financial support from the DFG for project SE 1119/3 is gratefully
acknowledged.

\end{document}